\documentstyle[aps,psfig,bbold]{revtex}

\newcommand{\beq}{\begin{equation}}
\newcommand{\eeq}{\end{equation}}
\newcommand{\beqa}{\begin{eqnarray}}
\newcommand{\eeqa}{\end{eqnarray}}

\begin{document}
\draft
\preprint{MKPH-T-01-10}

\title{
\hfill{\small {\bf MKPH-T-01-10}}\\
{\bf Construction of Consistent Meson Exchange Currents by Laplace 
Transform}\footnote[2]
{Supported by the Deutsche Forschungsgemeinschaft (SFB 443).}
}
\author
{Hartmuth Arenh\"ovel and Michael Schwamb}
 \address{ Institut f\"ur Kernphysik,           
  Johannes Gutenberg-Universit\"at,  
  D-55099 Mainz, Germany }
\date{\today}
\maketitle
\begin{abstract}
We propose a new method for the construction of a consistent meson exchange 
current in $r$-space for the spin-isospin dependent central and tensor part of 
phenomenological nucleon-nucleon potentials by using a Laplace transformation,
which allows the representation by a finite number of Yukawa functions. 
This method is applied to the Paris and the recent Argonne $V_{18}$ potentials.
Results are presented for electrodisintegration of the deuteron near threshold.
\end{abstract}


\section{Introduction}\label{intro}
The contribution of meson exchange currents (MEC) to electromagnetic reactions 
(e.m.)\ on nuclei, like photo induced reactions and electron scattering, 
constitute an 
important manifestation of meson degrees of freedom mediating the strong 
interactions between nucleons in nuclei. They describe the e.m.\ interaction 
with a nucleus during the interaction of nucleons and appear formally in 
the nuclear current operator as two- or many-body operators. Thus they are 
intimately linked to the $NN$-interaction. However, for a given $NN$-potential 
there exists no a priori way of constructing the appropriate MEC-operators, 
unless the $NN$-potential is derived from an underlying more fundamental 
fieldtheoretical 
framework with explicit subnuclear degrees of freedom which allows the 
construction of the corresponding two-body currents in parallel to the 
$NN$-potential. Although the existence of exchange currents associated with
an exchange $NN$-potential has been acknowledged for a long time, the early 
realistic potentials being phenomenological to a large extent prevented thus
the explicit consideration of such exchange currents. 

A breakthrough came with the meson-theory based construction of 
MEC-operators. Thus purely meson exchange models of the $NN$-interaction 
like the Bonn potentials~\cite{Mac89} allow one to build the appropriate 
consistent MEC-operators uniquely (for a recent derivation including leading 
order relativistic contributions see~\cite{RiG97}). In these studies 
it turned out that the most important MEC contributions came from the $\pi$- 
and $\rho$-MEC which are directly related to the spin-isospin dependent 
central and tensor part of the $NN$-potential~\cite{ChR71,Mat89,Ris89,Are99}. 
But for largely 
phenomenological potentials the construction of a consistent MEC remained
questionable. In view of the fact that realistic phenomenological potentials 
incorporated $\pi$-exchange as longest range contribution, one often used for
such potentials only a regularized $\pi$-MEC as an approximation. 

However, the above mentioned results for meson-theoretical models suggested
a method which provides also for the spin-isospin dependent 
central and tensor part of a phenomenological potential a consistent MEC
whose construction is based on an analogy with the properties of $\pi$- 
and $\rho$-MEC~\cite{BuL85,Ris85}. Essentially this method relies on a 
splitting of the spin-isospin dependent central and tensor part of a given 
$NN$-potential into a $\pi$- and $\rho$-like part for which the consistent
MECs are known. While the approach of Riska~\cite{Ris85} is based on the 
momentum space representation of the potential and can be applied to any given 
phenomenological potential, the method of Buchmann et al.~\cite{BuL85} is 
conceived for
$r$-space calculations and needs a representation of the potential as 
superposition of Yukawa functions. Thus the application of the latter method
seems to be more limited because not all phenomenological potentials have such
a form. On the other hand, in view of recent high precision, though largely
phenomenological $NN$ potentials, like the Argonne $V_{18}$~\cite{WiS95}, it
would be desirable to have a method available which allows the construction 
of a consistent exchange current for such potentials directly in $r$-space 
representation.

It is the aim of the present brief note to show, that it is possible to 
represent a given realistic $r$-space potential in general as a superposition
of Yukawa functions such that the construction of a consistent MEC in 
$r$-space 
is easily achieved. This new method is based on the representation of the 
radial central and tensor parts of a spin-isospin dependent potential by 
a Laplace transformation. First, we will briefly review in the next section 
the approaches of~\cite{BuL85,Ris85}. Sect.~\ref{laplace} contains the central 
idea introducing the Laplace transformation. Explicit applications to the 
Paris and Argonne potentials are presented in Sect.~\ref{potential}. In
Sect.~\ref{results} we consider electrodisintegration of the deuteron near
threshold as a test case for the evaluation of the corresponding MECs. 
Finally we close with a 
summary.

\section{Basic Formalism}\label{formal}
We will focus now on the spin-isospin dependent part of a given $NN$-potential
having the general form
\beq
V^{\sigma\tau}(r)= \vec \tau_1\cdot\vec\tau_2\,
\Big(\vec \sigma_1\cdot\vec\sigma_2\,V_C^{\sigma\tau}(r)+S_{12}\,
V_T^{\sigma\tau}(r)\Big)\,,\label{vsigtau}
\eeq
where
\beq
S_{12}=3\,\vec\sigma_1\cdot\hat r\,\vec\sigma_2\cdot\hat r
-\vec \sigma_1\cdot\vec\sigma_2
\eeq
denotes the usual spin tensor operator. A pure unregularized $\pi$-exchange 
potential is given by
\beq
V_\pi^{\sigma\tau}(\vec r\,)=\vec \tau_1\cdot\vec\tau_2\,
\frac{3\,V_{\pi}^0}{m_\pi^2}\,
\vec\sigma_1\cdot\vec\nabla \,\vec\sigma_2\cdot\vec\nabla\,J_{m_{\pi}}(r)\,,
\eeq
with the potential strength denoted by $V_{\pi}^0$ and
\beq
J_{m}(r)=\frac{e^{-mr}}{4\pi r}\,.
\eeq
By recoupling 
\beq
\vec\sigma_1\cdot\vec\nabla \,\vec\sigma_2\cdot\vec\nabla = 
\frac{1}{3}\Big(\vec \sigma_1\cdot\vec\sigma_2\, \partial^2_{r,C}
+S_{12}\partial^2_{r,T}\Big)
\eeq
with radial differentiation operators
\beq
\partial^2_{r,C}=\frac{1}{r}\frac{d^2}{dr^2}(r \,\cdot)\quad\mbox{and}\quad
\partial^2_{r,T}=r\,\frac{d}{dr}(\frac{1}{r}\,\frac{d}{dr} \,\cdot)\,,
\eeq
and using 
\beqa
\partial^2_{r,C}\,J_{m}(r)&=&m^2\,J_{m}(r)-\delta(\vec r\,)\,,\label{drcJ}\\
\partial^2_{r,T}\,J_{m}(r)&=&m^2\,F_T(mr)\,J_{m}(r)\,,\label{drtJ}
\eeqa
with
\beq
F_T(x)=1+\frac{3}{x}\,(1+\frac{1}{x})\,,
\eeq
it can be brought into the form (\ref{vsigtau}) with
\beqa
V_C^{\pi}(r)&=&\frac{V_{\pi}^0}{m_\pi^2}\,
\partial^2_{r,C}\,J_{m_{\pi}}(r)\nonumber\\
&=&V_{\pi}^0\Big(J_{m_{\pi}}(r)-\frac{1}{m_\pi^2}\,\delta(\vec r\,)\Big)\,,\\
V_T^{\pi}(r)&=&\frac{V_{\pi}^0}{m_\pi^2}\,
\partial^2_{r,T}\,J_{m_{\pi}}(r)\nonumber\\
&=&V_{\pi}^0\,F_T(m_{\pi}r)\,J_{m_{\pi}}(r)\,.
\eeqa
Analogously, the dominant part of the $\rho$-exchange potential has also 
the form (\ref{vsigtau}) with 
\beq
V_C^{\rho}(r)=2\,V_{\rho}^0\,\partial^2_{r,C}\,J_{m_{\rho}}(r)\quad\mbox{and}\quad
V_T^{\rho}(r)=-V_{\rho}^0\,\partial^2_{r,T}\,J_{m_{\rho}}(r)\,.
\eeq
The potential strength is denoted by $V_{\rho}^0$.

The corresponding exchange currents read for $\pi$-exchange 
\beqa
\vec \jmath_{\pi} (\vec x,\vec r_1, \vec r_2 )= T_{12}^3
\,V_{\pi}^0\,\Big(&&\delta(\vec x-\vec r_1)\,\vec\sigma_1
\,\vec\sigma_2\cdot\vec\nabla_2\,J_{m_{\pi}}(|r_1-r_2|) 
-(1\leftrightarrow 2)\nonumber\\
&&+\vec\sigma_1\cdot\vec\nabla_1\,\vec\sigma_2\cdot\vec\nabla_2\,
J_{m_{\pi}}(|\vec r_1 -\vec x|)\,\stackrel{\leftrightarrow}{\nabla}_x\,
J_{m_{\pi}}(|\vec x -\vec r_2|)\Big)\,,\label{pimec}
\eeqa
and for $\rho$-exchange
\beqa
\vec \jmath_{\rho} (\vec x,\vec r_1, \vec r_2 )= T_{12}^3
\,V_{\rho}^0\,\Big(&&\delta(\vec x-\vec r_1)\,
\,(\vec\sigma_2\times\vec\nabla_2)\times\vec\sigma_1\,J_{m_{\rho}}(|r_1-r_2|) 
-(1\leftrightarrow 2)\nonumber\\
&&+(\vec\sigma_1\times\vec\nabla_1)\cdot(\vec\sigma_2\times\vec\nabla_2)\,
J_{m_{\rho}}(|\vec r_1 -\vec x|)\,\stackrel{\leftrightarrow}{\nabla}_x\,
J_{m_{\rho}}(|\vec x -\vec r_2|)\Big)\,,\label{rhomec}
\eeqa
where we have introduced 
\beq
T_{12}^3=(\vec \tau_1\times\vec\tau_2)_3\,.
\eeq
Defining the currents in momentum space by
\beq
\vec \jmath\,(\vec q,\vec q_1, \vec q_2 )=\frac{1}{(2\pi)^3}\,
\int d^3x\,d^3r_1d^3r_2 e^{-i\vec q\cdot \vec x}\,e^{i\vec q_1\cdot \vec r_1}\,
e^{i\vec q_2\cdot \vec r_2}\,\vec \jmath\,(\vec x,\vec r_1, \vec r_2 )\,,
\eeq
these currents read
\beqa
\vec\jmath_{\pi}\,(\vec{q},\vec q_1, \vec q_2)&=&
-i\,\delta(\vec q-\vec q_1- \vec q_2)\,T_{12}^3\big[\vec{\sigma}_1\,
(\vec{\sigma}_2\,\cdot\,\vec{q}_2)\,
v_{\pi}(\vec{q}_2)-(1\leftrightarrow 2)\nonumber\\
&&+\frac{\vec{q}_1-\vec{q}_2}{q_1^2-q_2^2}\,(\vec{\sigma}_1\,\cdot\,\vec{q}_1)
(\vec{\sigma}_2\,\cdot\,\vec{q}_2)\,\big(v_{\pi}(\vec{q}_1)
-v_{\pi}(\vec{q}_2)\big)\big]\,,\label{pimecp}\\
\vec\jmath_{\rho}\,(\vec{q},\vec q_1, \vec q_2)&=&-i\,
\delta(\vec q-\vec q_1- \vec q_2)\,T_{12}^3\,\big[\vec{\sigma}_1\,\times
(\vec{\sigma}_2\times\vec{q}_2)\,
v_{\rho}(\vec{q}_2)-(1\leftrightarrow 2)\nonumber\\
&&+\frac{\vec{q}_1-\vec{q}_2}{q_1^2-q_2^2}\,(\vec{\sigma}_1\times\vec{q}_1)
\,\cdot\,(\vec{\sigma}_2\times\vec{q}_2)\,\big(v_{\rho}(\vec{q}_1)
-v_{\rho}(\vec{q}_2)\big)
\big]\,,\label{rhomecp}
\eeqa
where
\beqa
v_{\pi/\rho}(\vec{q}\,)&=&\int d^3r V_{\pi/\rho}(r)
e^{i\vec{q}\cdot\vec{r}}\,
\eeqa
denotes the corresponding Fourier transforms of the potentials.

As mentioned in the introduction, 
for the case of phenomenological central and tensor parts of a realistic 
spin-isospin dependent $NN$-potential, the construction of a consistent MEC 
is based on the idea to split these potential terms into a $\pi$-like and a 
$\rho$-like part for which the corresponding MEC are known. This method has 
been developed independently by Buchmann et al.~\cite{BuL85} and 
Riska~\cite{Ris85} for the Paris potential~\cite{LaL80}. Buchmann et al.\ 
start from the representation of the Paris potential in terms of Yukawa 
functions which reads in particular for the spin-isospin dependent part 
\beqa
V_C^{\sigma\tau}(r)&=&\sum_{j=1}^{12} g_{C,j}^{\sigma\tau}\,J_{m_j}(r)\,,
\label{vparisC}\\
V_T^{\sigma\tau}(r)&=&\sum_{j=1}^{12} g_{T,j}^{\sigma\tau}\,
F_T(m_j r)\,J_{m_j}(r)\,.
\label{vparisT}
\eeqa
Here $m_j$, $g_C^{\sigma\tau}$ and $g_C^{\sigma\tau}$ denote appropriate 
masses and coupling constants (for details see~\cite{BuL85}). Using 
(\ref{drcJ}) and (\ref{drtJ}) one finds
\beqa
V_C^{\sigma\tau}(r)&=&\partial^2_{r,C} \sum_{j=1}^{12} 
\frac{g_{C,j}^{\sigma\tau}}{m_j^2}\,J_{m_j}(r) + 
\Big(\sum_{j=1}^{12} \frac{g_{C,j}^{\sigma\tau}}{m_j^2}\Big)\delta(\vec r\,)
\,,\\
V_T^{\sigma\tau}(r)&=&\partial^2_{r,T} \sum_{j=1}^{12} 
\frac{g_{T,j}^{\sigma\tau}}{m_j^2}\,J_{m_j}(r)
\eeqa
These expressions can be rewritten as
\beqa
V_C^{\sigma\tau}(r)&=&\partial^2_{r,C}
\Big(V^{\pi-like}(r)+2\,V^{\rho-like}(r)\Big)+
\Big(\sum_{j=1}^{12} \frac{g_{C,j}^{\sigma\tau}}{m_j^2}\Big)\delta(\vec r\,)
\,,\label{pariscdelta}\\
V_T^{\sigma\tau}(r)&=&\partial^2_{r,T}
\Big(V^{\pi-like}(r)-V^{\rho-like}(r)\Big)\,,\label{paristdelta}
\eeqa
where the $\pi$- and $\rho$-like parts are given by 
\beq
V^{(\pi/\rho)-like}(r)=\sum_{j=1}^{12} g_{j}^{\pi/\rho}
\,J_{m_j}(r)\,,
\eeq
with coupling constants defined by
\beq
g^{\pi}_j=\frac{1}{3m_j^2}\,(g_{C,j}^{\sigma\tau}+2\,g_{T,j}^{\sigma\tau})
\quad\mbox{and}\quad
g^{\rho}_j=\frac{1}{3m_j^2}\,(g_{C,j}^{\sigma\tau}-g_{T,j}^{\sigma\tau})\,.
\eeq
Thus the representation of the spin-isospin dependent part of the Paris 
potential in terms of $\pi$- and $\rho$-like potentials is achieved if 
the $\delta$-function disappears in (\ref{pariscdelta}), which means 
the coupling constants have to fulfil the following condition
\beq
\sum_{j=1}^{12}(g^{\pi}_j+2\,g^{\rho}_j)=
\sum_{j=1}^{12}\frac{g_{C,j}^{\sigma\tau}}{m_j^2}=0\,.\label{delta_cond}
\eeq
If this condition is not fulfilled, as is the case for the Paris potential, 
one can modify in (\ref{vparisC}) the short range part of 
$V_C^{\sigma\tau}(r)$ by splitting it into
\beq
V_C^{\sigma\tau}(r)= \widetilde V_C^{\sigma\tau}(r) + V_C^{sr}(r)
\label{shortrange}
\eeq
such that $\widetilde V_C^{\sigma\tau}(r)$ is identical to 
$V_C^{\sigma\tau}(r)$ in the long and medium range part but obeys 
(\ref{delta_cond}). In \cite{BuL85} this is achieved by changing the 
coupling constant of the highest mass $g_{C,12}^{\sigma\tau}\rightarrow \tilde 
g_{C,12}^{\sigma\tau}$, where the latter is determined from (\ref{delta_cond}),
i.e.\ from
\beq
\tilde g_{C,12}^{\sigma\tau}=-m_{12}^2\,\sum_{j=1}^{11}
\frac{g_{C,j}^{\sigma\tau}}{m_j^2}\,.
\eeq
The difference $g_{C,sr}^{\sigma\tau}=g_{C,12}^{\sigma\tau}-\tilde 
g_{C,12}^{\sigma\tau}$ serves as coupling constant for 
$V_C^{sr}(r)= g_{C,sr}^{\sigma\tau}\,J_{m_{12}}(r)$. For the remaining 
short range potential $V_C^{sr}$ a consistent MEC is easily 
constructed~\cite{BuL85}.
It is obvious that this procedure is not unique and that it introduces some
ambiguity, which indeed is characteristic for such phenomenological 
approaches. The hope is that the important physics in the long and medium 
range part is preserved. 

Riska on the other hand considers the momentum space representation of
the potential terms obtaining the $\pi$- and $\rho$-like pieces from
\beqa
v^{\pi-like}(p)&=&\frac{1}{3p^2}\Big[4\pi\,\int dr r^2\Big(V_C^{\sigma\tau}(r)
j_0(pr)+2\,V_T^{\sigma\tau}(r)j_2(pr)\Big) - v_C^{\sigma\tau}(0)\Big]\,,\\
v^{\rho-like}(p)&=&\frac{1}{3p^2}\Big[4\pi\,\int dr r^2\Big(V_C^{\sigma\tau}(r)
j_0(pr)-V_T^{\sigma\tau}(r)j_2(pr)\Big) - v_C^{\sigma\tau}(0)\Big]\,,
\eeqa
where the subtraction of the term 
\beq
v_C^{\sigma\tau}(0)= 4\pi\,\int dr r^2V_C^{\sigma\tau}(r)
\eeq
is required in order to eliminate the $\delta$-function. It 
constitutes again a modification of the short range part of the original 
potential part as mentioned above. In fact, in this case it amounts to the
subtraction of a $\delta$-function
\beq
V_C^{sr}(\vec r\,)= v_C^{\sigma\tau}(0)\,\delta(\vec r\,)\,.
\eeq
Also here the above mentioned ambiguity becomes apparent, because any 
finite range potential $V_C^{sr}(r)$ with the property
\beq
4\pi\,\int dr r^2\,V_C^{sr}(r)= v_C^{\sigma\tau}(0)
\eeq
would also serve to eliminate the $\delta$-function. 

The corresponding MECs are obtained from (\ref{pimec}) and (\ref{rhomec})
by replacing the Fourier transforms $v_\pi$ and $v_\rho$ by the 
corresponding ones, $v^{\pi-like}$ and $v^{\rho-like}$, 
respectively. This method appears more general than the approach of Buchmann 
et al.\ since it does not rely on the Yukawa representation. However, we 
now will show that also the latter method can be applied to potentials 
with a more general radial dependence.

\section{Representation by a Laplace Transform}\label{laplace}
Our first idea was to approximate the given potentials by a series of 
Yukawa functions whose coefficients and masses are obtained by a least 
square fit. But depending on the accuracy needed this can be quite a 
formidable task because of the high dimensional parameter space involved. 
But then it turned out that an easier and more systematic approach
can be based on the representation of the potentials by a Laplace 
transform. Indeed, because of the fact that realistic potentials contain as 
longest range contribution a $\pi$-exchange potential, one can represent 
the central and tensor parts of a given realistic potential by a continuous 
superposition of appropriate Yukawa functions (compare with (\ref{vparisC}) 
and  (\ref{vparisT})), i.e.\
\beqa
V_C(r)&=& \int_0^\infty dm\,g_C(m)\,J_{m+m_\pi}(r)\nonumber\\
&=& J_{m_\pi}(r)\,\int_0^\infty dm\,g_C(m)\,e^{-m r}\,,
\label{vclaplace}\\
V_T(r)&=& \int_0^\infty dm\,g_T(m)\,F_T((m+m_\pi)r)\,
J_{m+m_\pi}(r)\nonumber\\
&=&\partial^2_{r,T}\,J_{m_\pi}(r)\,
\int_0^\infty dm\,\frac{g_C(m)}{(m+m_\pi)^2}\,e^{-m r}
\,,\label{vtlaplace}
\eeqa
which is essentially a Laplace transform representation. We note in passing, 
that the Fourier transform of $V_C$ is simply given by
\beqa
v_C(\vec q\,)&=&\int_0^\infty dm\,\frac{g_C(m)}{\vec q^{\,2}+(m+m_\pi)^2}\,.
\eeqa

Introducing $\pi$- and $\rho$-like potentials as in (\ref{pariscdelta}) and 
(\ref{paristdelta}) with corresponding representations 
\beqa
V^{\pi/\rho-like}(r)&=& \int_0^\infty dm\,g_{\pi/\rho}(m)\,J_{m+m_\pi}(r)\,,
\eeqa
one obtains the following relations
\beqa
V_C(r)&=&\int_0^\infty dm\,(g_\pi(m)+2\,g_\rho(m))\,
\Big((m+m_\pi)^2\,J_{m+m_\pi}(r)-\delta(\vec r\,)\Big)
\,,\label{vclaplace1}\\
V_T(r)&=&\int_0^\infty dm\,(m+m_\pi)^2(g_\pi(m)-g_\rho(m))\,
F_T((m+m_\pi)r)\,J_{m+m_\pi}(r)\,.\label{vtlaplace1}
\eeqa
Comparison with (\ref{vclaplace}) and (\ref{vtlaplace}) gives
\beqa
g_C(m)&=&(m+m_\pi)^2(g_\pi(m)+2\,g_\rho(m))\,,\\
g_T(m)&=&(m+m_\pi)^2(g_\pi(m)-g_\rho(m))\,.
\eeqa
Again, in order to eliminate the $\delta$-function in (\ref{vclaplace1}),
one needs the condition
\beq
\int_0^\infty dm\,(g_\pi(m)+2\,g_\rho(m))=
\int_0^\infty dm\,\frac{g_C(m)}{(m+m_\pi)^2}=0\,.\label{g_j_condition}
\eeq
If this is not fulfilled, one has to separate again a short range potential
as in (\ref{shortrange})
\beq
V_C^{sr}(r)=\frac{1}{r}\,\int_0^\infty dm\,g_C^{sr}(m)\,e^{-(m+m_\pi)r}\,,
\eeq
where the coefficients $g_C^{sr}(m)$ in principle 
can be chosen quite arbitrarily except for the fulfilment of the relation
\beq
\int_0^\infty dm\,\frac{g_C^{sr}(m)}{(m+m_\pi)^2}=
\int_0^\infty dm\,\frac{g_C(m)}{(m+m_\pi)^2}\,.
\eeq
However, in practice one would choose them such that only the short-range part 
of the original potential is modified. The associated exchange currents read
\beqa
\vec \jmath_{\pi-like} (\vec x,\vec r_1, \vec r_2 )= T_{12}^3
\,\int_0^\infty dm\,g_\pi(m)\,\Big(&&\delta(\vec x-\vec r_1)\,\vec\sigma_1
\,\vec\sigma_2\cdot\vec\nabla_2\,J_{m+m_{\pi}}(|r_1-r_2|) 
-(1\leftrightarrow 2)\nonumber\\
&&+\vec\sigma_1\cdot\vec\nabla_1\,\vec\sigma_2\cdot\vec\nabla_2\,
J_{m+m_{\pi}}(|\vec r_1 -\vec x|)\,\stackrel{\leftrightarrow}{\nabla}_x\,
J_{m+m_{\pi}}(|\vec x -\vec r_2|)\Big)\,,\label{pimeclike}\\
\vec \jmath_{\rho-like} (\vec x,\vec r_1, \vec r_2 )= T_{12}^3
\,\int_0^\infty dm\,g_\rho(m)\,\Big(&&\delta(\vec x-\vec r_1)\,
\,(\vec\sigma_2\times\vec\nabla_2)\times\vec\sigma_1\,J_{m+m_{\pi}}(|r_1-r_2|) 
-(1\leftrightarrow 2)\nonumber\\
&&+(\vec\sigma_1\times\vec\nabla_1)\cdot(\vec\sigma_2\times\vec\nabla_2)\,
J_{m+m_{\pi}}(|\vec r_1 -\vec x|)\,\stackrel{\leftrightarrow}{\nabla}_x\,
J_{m+m_{\pi}}(|\vec x -\vec r_2|)\Big)\,.\label{rhomeclike}
\eeqa

For the explicit application it is useful to discretize the integrals in 
(\ref{vclaplace}) and (\ref{vtlaplace}), for example by Gauss quadrature
\beq
\int_0^\infty dm\,h(m)=\sum_{j=1}^N w_j\,h(m_j)\,,
\eeq
where $N$ denotes the number of Gauss points, $m_j$ the Gauss points and 
$w_j$ the corresponding weights. Then one can represent the radial functions
in (\ref{vclaplace1}) and (\ref{vtlaplace1}) by a finite number of Yukawa 
functions 
\beqa
V_C(r)&=& \sum_{j=1}^N w_j\,(g_\pi(m_j)+2\,g_\rho(m_j))\,
\mu_j^2\,J_{\mu_j}(r)\,,\label{vclaplace2}\\
V_T(r)&=& \sum_{j=1}^N w_j\,\mu_j^2
(g_\pi(m_j)-g_\rho(m_j))\,F_T(\mu_jr)\,J_{\mu_j}(r)\,,
\eeqa
where we have set $\mu_j=m_j+m_\pi$. The consistent exchange current is 
then given as superposition of $\pi$- and $\rho$-like currents as listed
in (\ref{pimec}) and (\ref{rhomec})
\beqa
\vec \jmath_{mec}&=&\vec \jmath_{\pi-like} (\vec x,\vec r_1, \vec r_2 )
+\vec \jmath_{\rho-like} (\vec x,\vec r_1, \vec r_2 )
\eeqa
with
\beqa
\vec \jmath_{\pi-like} (\vec x,\vec r_1, \vec r_2 )= T_{12}^3
\,\sum_{j=1}^N w_j\,g_\pi(m_j)&\Big(&\delta(\vec x-\vec r_1)\,\vec\sigma_1
\,\vec\sigma_2\cdot\vec\nabla_2\,J_{\mu_j}(|r_1-r_2|) 
-(1\leftrightarrow 2)\nonumber\\
&&+\vec\sigma_1\cdot\vec\nabla_1\,\vec\sigma_2\cdot\vec\nabla_2\,
J_{\mu_j}(|\vec r_1 -\vec x|)\,\stackrel{\leftrightarrow}{\nabla}_x\,
J_{\mu_j}(|\vec x -\vec r_2|)\Big)\,,\label{pilikemec}\\
\vec \jmath_{\rho-like} (\vec x,\vec r_1, \vec r_2 )= T_{12}^3
\,\sum_{j=1}^N w_j\,g_\rho(m_j)&\Big(&\delta(\vec x-\vec r_1)\,
\,(\vec\sigma_2\times\vec\nabla_2)\times\vec\sigma_1\,J_{\mu_j}(|r_1-r_2|) 
-(1\leftrightarrow 2)\nonumber\\
&&+(\vec\sigma_1\times\vec\nabla_1)\cdot(\vec\sigma_2\times\vec\nabla_2)\,
J_{\mu_j}(|\vec r_1 -\vec x|)\,\stackrel{\leftrightarrow}{\nabla}_x\,
J_{\mu_j}(|\vec x -\vec r_2|)\Big)\,.\label{rholikemec}
\eeqa

In order to determine the unknown coefficients, one can choose an
appropriate grid of $N$ radial points $r_i$ ($i=1,\dots, N$), and obtains 
an inhomogeneous set of $2N$ linear equations for the $2N$ coefficients 
$g_\pi(m_j)$ and $g_\rho(m_j)$
\beqa
c_i&=&\sum_{j=1}^N\,A^C_{ij}\,(\pi_j+2\,\rho_j)\label{lineqc}\,,\\
t_i&=&\sum_{j=1}^N\,A^T_{ij}\,(\pi_j-\rho_j)\label{lineqt}\,,
\eeqa
where we have introduced for convenience
\beq
\begin{array}{rclrcl}
\pi_j&=&g_\pi(m_j)\,,& \quad\rho_j&=&g_\rho(m_j),,\\
c_i&=&{V_C(r_i)}/{J_{m_\pi}(r_i)}\,,
&\quad t_i&=&{V_T(r_i)}/{J_{m_\pi}(r_i)}\,,\\
A^C_{ij}&=&w_j\,\mu_j^2\,e^{-m_j r_i}\,,&\quad
A^T_{ij}&=&w_j\,\mu_j^2\,F_T(\mu_jr_i)\,e^{-m_j r_i}\,.\\
\end{array}
\eeq
It is worth mentioning that the matrices $A^C_{ij}$ and $A^T_{ij}$ do not 
depend on the potentials, only on the chosen grids of radial points $r_i$ 
and masses $m_j$. 

Solving the equations (\ref{lineqc}) and (\ref{lineqt}) by matrix inversion, 
one finds for the coefficients
\beqa
\pi_j&=&\frac{1}{3}\,\sum_{i=1}^N 
\Big( (A^C)^{-1}_{ji}\,c_i+2\,(A^T)^{-1}_{ji}\,t_i\Big)\,,\label{pi_j}\\
\rho_j&=&\frac{1}{3}\,\sum_{i=1}^N 
\Big( (A^C)^{-1}_{ji}\,c_i-(A^T)^{-1}_{ji}\,t_i\Big)\,.\label{rho_j}
\eeqa
The condition (\ref{g_j_condition}) becomes
\beq
\sum_{j=1}^N w_j\,(\pi_j+2\,\rho_j)=0\,.\label{g_j_cond}
\eeq
If this is not obeyed for the given central potential, we split off 
a short range Yukawa potential for the highest mass $\mu_N=m_N+m_\pi$
\beq
V_C^{sr}(r)=c \,\frac{e^{-\mu_N r}}{r}\,.
\eeq
This means that the coefficients $c_i$ in (\ref{pi_j}) and (\ref{rho_j}) 
have to be changed according to 
\beq
c_i \rightarrow c_i +c\,e^{-m_N r_i}\,,
\eeq
leading to new coefficients $\pi_j$ and $\rho_j$.
Then the unknown coefficient $c$ is determined 
by the requirement that the resulting new coefficients $\pi_j$ 
and $\rho_j$ fulfil (\ref{g_j_cond}), which leads to
\beq
c=-\frac{\sum_{i,j=1}^N w_j\,(A^C)^{-1}_{ji}\,c_i}
{\sum_{i,j=1}^N w_j\,(A^C)^{-1}_{ji}\,e^{-m_N r_i}}\,.
\eeq

\section{Application to Paris and Argonne $V_{18}$ Potentials}\label{potential}

As first example, we have chosen the Paris potential~\cite{LaL80}, 
because in this case we can compare the Laplace transform representation 
of the associated MECs directly with the consistent $\pi$- and $\rho$-like 
MECs of~\cite{BuL85}. Since for a numerical evaluation one needs a reliable 
representation only for the radial range $r=0- 10$~fm, we have chosen 
the grid of $N$ radial points $r_j$ in the range between 0 and about 12~fm 
with variable 
stepsize with the highest density of points close to the origin, 
where the potentials exhibit the largest variation, and then with increasing 
stepsize approaching the highest value $r_N$. In fact, an educated choice 
will have to take into account the radial behaviour of the potential under 
consideration. In detail, we have chosen 
the $r$-grid to be defined by the following expression
\beq
r_j=r_0+(e^{a(j-1)}-1)\,e^{b(j-1)}\,\,\mbox{fm}\,,
\eeq
with $r_0=0.01$~fm, $a=0.01$, and $b$ is determined for a given $N$ 
by the requirement that the highest point $r_N$ lies approximately between
10 and 12~fm. The parameter $b$ is listed in Tab.~\ref{tab1}. In order to 
check the convergence with respect to the number of points we have considered 
$N=12,\dots,20$ in steps of 2. The Gauss points and weights for the integral 
over $m$ have been chosen according to
\beqa
m_j&=&s\,\tan(\frac{\pi}{4}\,x_j+1)\,,\\
w_j&=&\frac{\pi}{4}\,\frac{s\,y_j}{\cos^2(\frac{\pi}{4}\,x_j+1)}\,,
\eeqa
where $x_j$ and $y_j$ are Gauss points and weights, respectively, for 
integration between $-1$ and 1. For a given $N$ the scale factor $s$ is 
determined by minimizing the mean absolute deviation between the 
Yukawa representation $V^{(Y,N)}$ and the original potential $V$ 
\beq
\Delta(V^{(Y,N)})= \frac{1}{r_{max}-r_0}
\int_{r_0}^{r_{max}}dr\,|V^{(Y,N)}(r)-V(r)|\,,
\eeq
evaluated between $r_0=.01$~fm and $r_{max}=10$~fm. The resulting 
scale factors and the relative mean deviations, i.e., 
$\Delta(V^{(Y,N)})$ divided by the average potential strength $|\bar V|$ 
\beq
|\bar V| = \frac{1}{r_{max}-r_0}\,\int_{r_0}^{r_{max}} dr\, |V(r)|\,,
\eeq
for the central and tensor potentials are also listed in Tab.~\ref{tab1}. 

\begin{figure}
\centerline{\psfig{figure=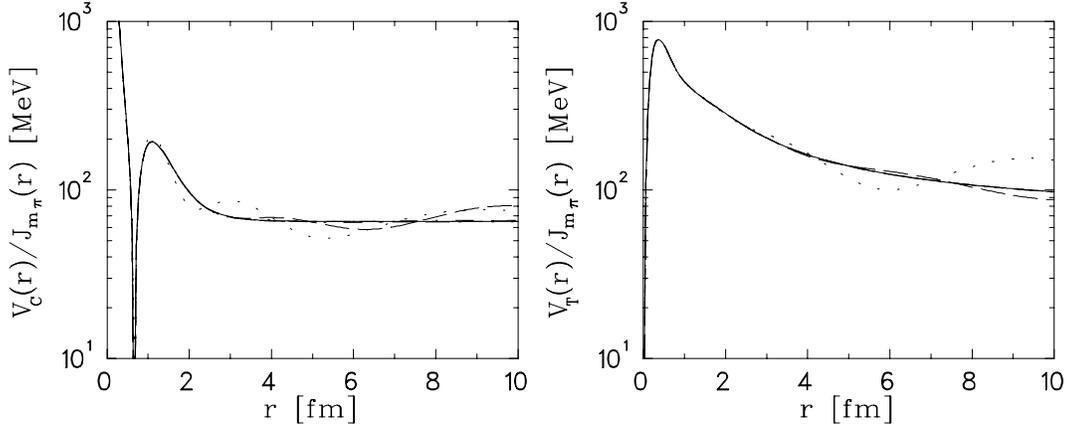,width=14cm,angle=0}}
\vspace{.5cm}
\caption{
Central and tensor spin-isospin dependent parts of Paris potential. Notation:
original form: full curves, Laplace transform representation for different
number $N$ of Gauss points: $N=12$: dotted, $N=14$: dashed, $N=16$: dash-dot.
}
\label{fig_vparis}
\end{figure}

\begin{figure}
\centerline{\psfig{figure=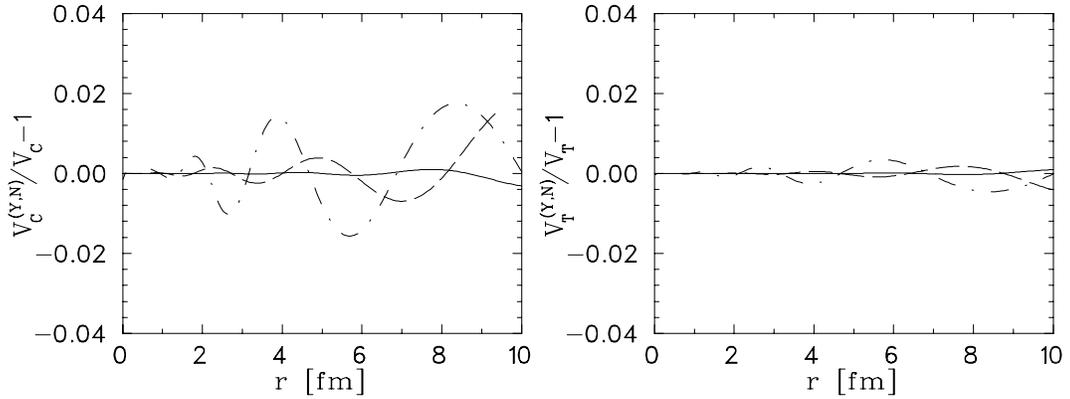,width=14cm,angle=0}}
\vspace{.5cm}
\caption{
Relative deviation of Laplace transform representation of central and tensor 
spin-isospin dependent parts of Paris potential from original form for 
different number $N$ of Gauss points: Notation: $N=16$: dash-dot,
$N=18$: long dashed, $N=20$: full curves.
}
\label{fig_paris_rel}
\end{figure}

In Fig.~\ref{fig_vparis} we show 
the original central and tensor potentials together with their Yukawa 
representations where we have 
multiplied them by the inverse of the pion Yukawa 
function $J_{m_\pi}(r)$ for $N=12,\dots,20$ in order to exhibit in greater 
detail the accuracy of the representation. 
One readily notes the rapid convergence and the very good representation 
over the whole range for $N\ge 16$. Indeed, the Laplace transform
representations for $N=18$ and 20 are undistinguishable from the
original form on this scale. Therefore, we show in addition in 
Fig.~\ref{fig_paris_rel} for $N=16,\,18$, and 20 the relative deviations of 
the Laplace transform representation from the original form. 
For small $r$-values up to about 3~fm the deviations are extremely small 
whereas for higher $r$-values the relative deviations 
become somewhat larger, but this is of little importance in view of the rapid 
fall-off of the potentials themselves with increasing $r$. It is
interesting to note that the relative deviations are larger for the
central part than for the tensor potential. The reason for this
feature is the rather rapid variation of the central potential near the origin.

\begin{figure}
\centerline{\psfig{figure=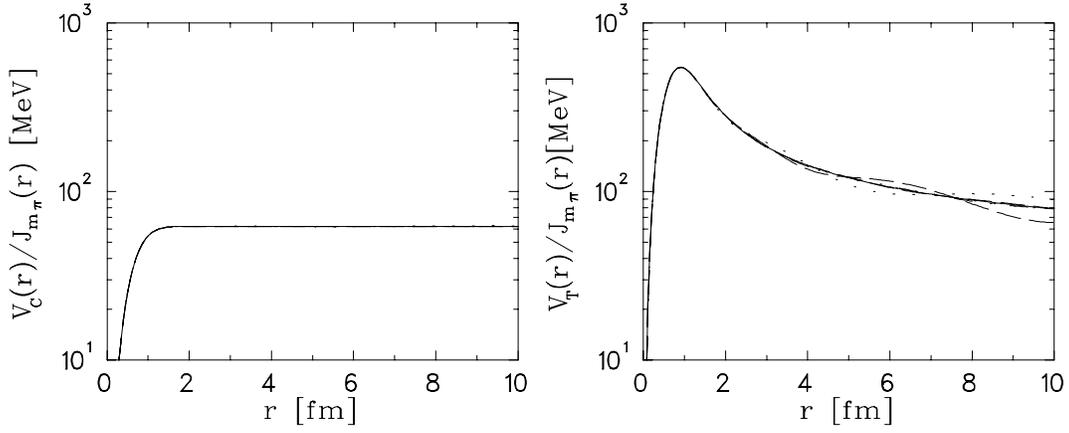,width=14cm,angle=0}}
\vspace{.5cm}
\caption{
Central and tensor spin-isospin dependent parts of Argonne $V_{18}$
potential: original form and Laplace transform representation. 
Notation as in Fig.~\ref{fig_vparis}. In the central part the various 
Laplace transform representations are almost undistinguishable from
original form. 
}
\label{fig_v18}
\end{figure}

\begin{figure}
\centerline{\psfig{figure=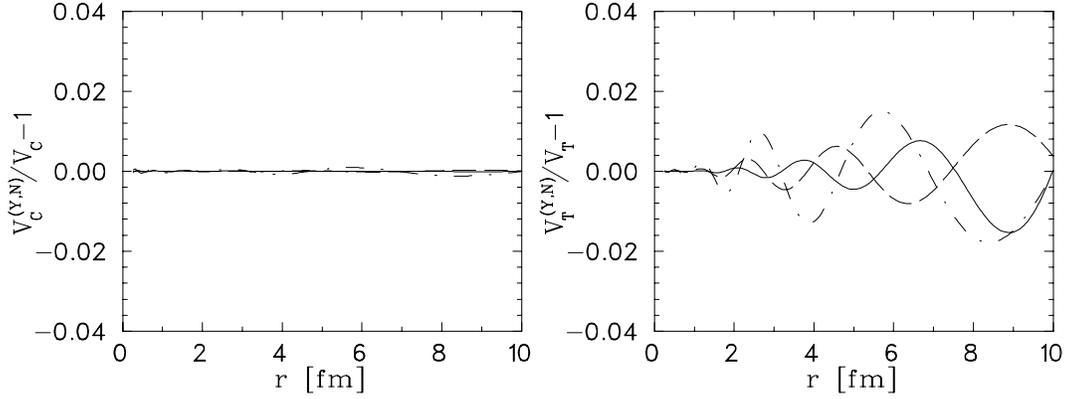,width=14cm,angle=0}}
\vspace{.5cm}
\caption{
Relative deviation of Laplace transform representation of central and 
tensor spin-isospin dependent parts of Argonne $V_{18}$ potential from 
original form for different number $N$ of Gauss points: Notation as in 
Fig.~\ref{fig_paris_rel}.
}
\label{fig_v18_rel}
\end{figure}

The analogous results for the Argonne $V_{18}$ potentials are displayed in 
Fig.~\ref{fig_v18} and Fig.~\ref{fig_v18_rel} with scale parameters and 
relative mean deviations listed in Table~\ref{tab2}. One readily notes again 
the excellent representation for $N=12,\dots,20$. In this case the
relative deviations are larger in the tensor part, because the central
potential exhibits a much smoother behaviour near the origin compared
to the Paris potential. In fact, the Laplace transform representations 
of the central part for $N=18$ and 20 are undistinguishable from the original 
form in Fig.~\ref{fig_v18}, and even on the magnifying scale of 
Fig.~\ref{fig_v18_rel} one notes only very tiny deviations. 

\section{Results for Deuteron Electrodisintegration near Threshold}
\label{results}

A benchmark for the study of meson exchange current effects in 
electromagnetic reactions on nuclei is deuteron electrodisintegration
near threshold at higher momentum
transfers~\cite{RiG97,HoR73,LoF75,FaA76,MoR76}. The threshold region is
dominated by the $M1$-excitation of the antibound $^1S_0$-resonance in
$NN$-scattering at very low energies. With increasing momentum
transfer the inclusive cross section at backward angles, where the
transverse current contribution via the inelastic transverse form factor 
dominates, the one-body current
contribution drops rapidly due a destructive interference of deuteron
$S$- and $D$-wave contributions. In this situation the contribution 
of MEC, which are of shorter range than the one-body currents, becomes 
relatively more important, in fact dominant. Only inclusion of such MEC 
gives a satisfactory description of experimental data yielding thus 
clear-cut evidence for the presence of exchange 
currents~\cite{Mat89,Ris89,Are99,HoR73,LoF75,FaA76,MoR76}. 

\begin{figure}
\centerline{\psfig{figure=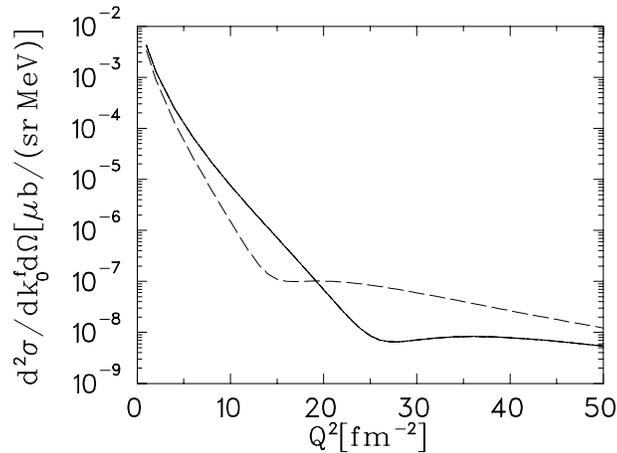,width=8cm,angle=0}}
\vspace{.5cm}
\caption{
Inclusive cross section for deuteron electrodisintegration near
threshold for Paris potential for final state excitation energy 
$E_{np}=1.5$~MeV and electron scattering angle
$\theta_e=155^\circ$. Notation: solid curve: consistent MEC of
\protect\cite{BuL85} coinciding with consistent MEC of Laplace transform 
representations for $N=12$, 16 and 20; dashed curve: without MEC. 
}
\label{fig_dees_paris}
\end{figure}

\begin{figure}
\centerline{\psfig{figure=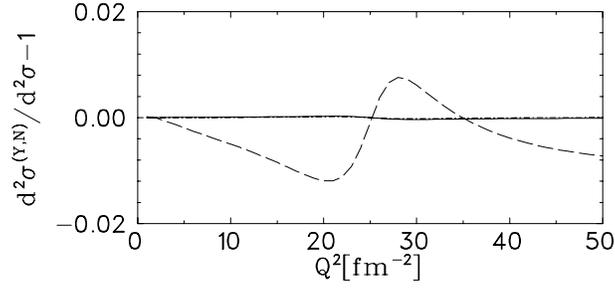,width=8cm,angle=0}}
\vspace{.5cm}
\caption{
Relative deviation 
of inclusive cross sections for deuteron electrodisintegration near 
threshold for Paris potential calculated with MEC from the Laplace
transform representation ($d^2\sigma^{(Y,N)}$) ($N=12,\,16,\,20$) to 
the one with MEC from~\protect\cite{BuL85} ($d^2\sigma$) for 
final state excitation energy $E_{np}=1.5$~MeV and electron scattering angle 
$\theta_e=155^\circ$. Notation of curves: $N=12$: dashed,
$N=16$: dash-dot, $N=20$: solid. 
}
\label{fig_dees_paris_rel}
\end{figure}

We show in Fig.~\ref{fig_dees_paris} the backward inclusive cross
section near threshold at a c.m.\ excitation energy of
$E_{np}=1.5$~MeV as function of the momentum transfer squared,
calculated for the consistent $\pi$- and $\rho$-like MEC according 
to~\cite{BuL85} and for the new Laplace transform representation for 
$N=12,\,16$ and 20. It is almost impossible to note a difference between 
the different curves on this scale. For this reason we exhibit in 
Fig.~\ref{fig_dees_paris_rel}
the relative deviation between the Laplace transform representation 
for $N=12,\,16$ and 20 and the original MEC of \cite{BuL85}. Already
for $N=12$ the maximum deviation does not exceed 1~\%, for $N=16$ it
is less than 0.1~\%. In fact, the difference to the $N=20$ result is hardly
noticeable even on the enlarged scale of Fig.~\ref{fig_dees_paris_rel}. 
For $N=20$ the agreement is perfect. This
clearly demonstrates that one has achieved already with $N=12$ quite a 
satisfactory parametrization, while for $N=16$ an almost perfect 
description for the consistent MEC is obtained. Here, we do not compare 
to experimental data for which one would need to include additional 
contributions, because we only want to demonstrate that the new 
method works very well. 

\begin{figure}
\centerline{\psfig{figure=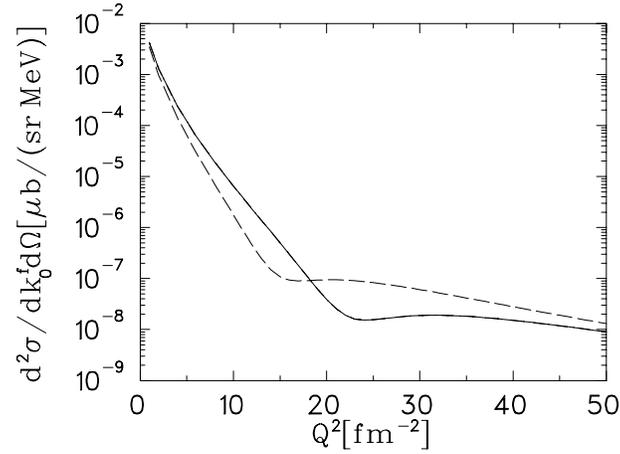,width=8cm,angle=0}}
\vspace{.5cm}
\caption{
Inclusive cross section for deuteron electrodisintegration near 
threshold for $V_{18}$ potential for final state excitation energy
$E_{np}=1.5$~MeV and electron scattering angle
$\theta_e=155^\circ$. Notation: dashed curve: without MEC; solid curve: 
consistent MEC of Laplace transform representations for $N=12$, 16 and 20,
which coincide with each other.
}
\label{fig_dees_v18}
\end{figure}

\begin{figure}
\centerline{\psfig{figure=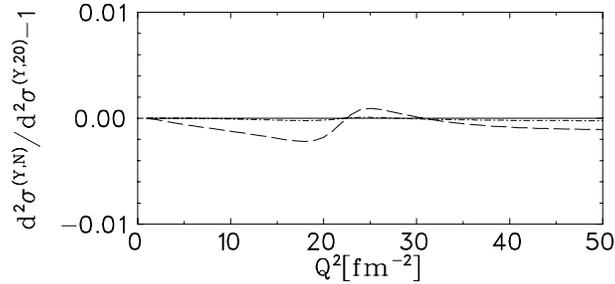,width=8cm,angle=0}}
\vspace{.5cm}
\caption{
Relative deviation 
of inclusive cross sections for deuteron electrodisintegration near 
threshold for $V_{18}$ potential calculated with MEC from the Yukawa 
representation ($d^2\sigma^{(Y,N)}$) for $N=12$ and 16 to the one with 
$N=20$ for final state excitation energy $E_{np}=1.5$~MeV and electron 
scattering angle $\theta_e=155^\circ$. Notation as in 
Fig.~\ref{fig_dees_paris_rel}.
}
\label{fig_dees_v18_rel}
\end{figure}

We then have evaluated the analogous MEC contributions for the Argonne
$V_{18}$ potential. Also in this case we found an excellent convergence 
of the Laplace transform representation as is demonstrated in 
Figs.~\ref{fig_dees_v18} and \ref{fig_dees_v18_rel} (note the further 
enlarged scale), where we show the same quantities obtained for
the Argonne $V_{18}$ potential as for the Paris potential. One readily
notes that in this case the convergence is even more rapid. For $N=12$ 
the maximum deviation from the case $N=20$ is about 0.2~\%, and 
for $N=16$ it is less than 0.01~\%.

\begin{figure}
\centerline{\psfig{figure=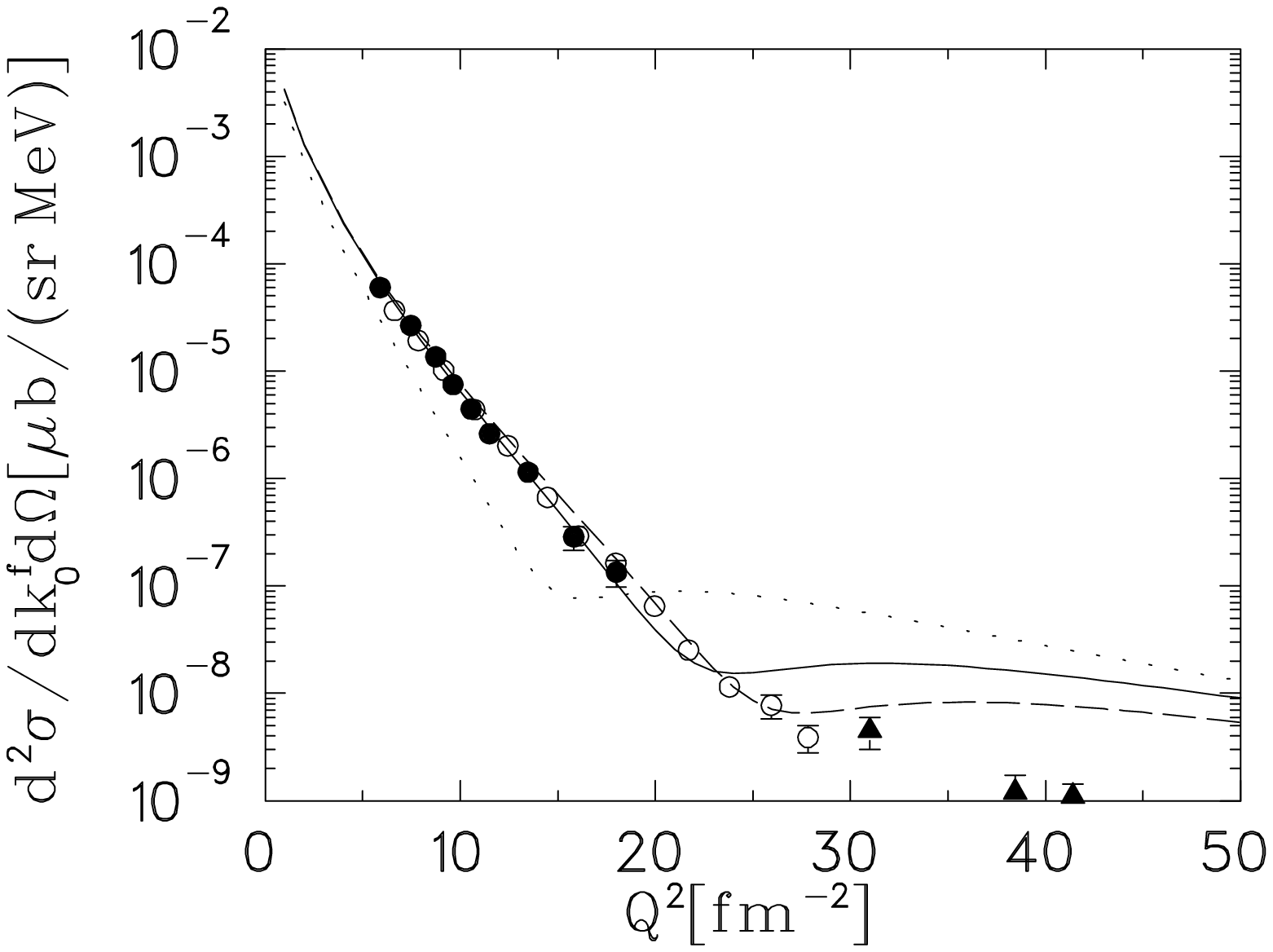,width=8cm,angle=0}}
\vspace{.5cm}
\caption{
Inclusive cross section for deuteron electrodisintegration near 
threshold for Paris and Argonne $V_{18}$ potentials for final state 
excitation energy
$E_{np}=1.5$~MeV and electron scattering angle
$\theta_e=155^\circ$. Notation: dotted curve: without MEC; solid curve: 
consistent MEC for Argonne $V_{18}$ 
potential; dashed curve: consistent MEC for Paris potential; 
Experiment: filled circles: \protect\cite{BeJ81}, open circles:
\protect\cite{Auf85} ($\theta_e=155^\circ$, averaged over energies
$0\mbox{\,MeV}\leq E_{np}\leq 3\mbox{\,MeV}$); filled triangles: 
\protect\cite{Arn90}
($\theta_e=180^\circ$, averaged over energies $0\mbox{\,MeV}\leq
E_{np}\leq 10\mbox{\,MeV}$).
}
\label{schwelle_vergleich_exp}
\end{figure}

We will end this section with a comparison with experimental data. 
In Fig.~\ref{schwelle_vergleich_exp} the theoretical results obtained for
the Paris and the Argonne potentials are exhibited together with 
experimental data from Refs.~\cite{BeJ81,Auf85,Arn90}. The theory 
includes besides the consistent $\pi$- and $\rho$-like MEC in 
addition relativistic one-body current and wave function boost contributions 
of leading order in $p/M$. One readily notes a satisfactory agreement for 
both potential models with
experiment up to a squared momentum transfer of about 25~fm$^{-2}$. At higher
momentum transfers the theory deviates significantly from experiment. 
However, in this region one expects a break down of the present approach
in view of the applied $p/M$-expansion~\cite{BeW94}. 

\section{Summary and Outlook}\label{summary}

In this note we have shown that it is possible to construct directly
in $r$-space a consistent meson exchange current for a spin-isospin
dependent $NN$-potential by representing the potential as a continuous
superposition of Yukawa functions, essentially a Laplace transform
representation. In this way it is possible to rewrite the potential
into a $\pi$- and a $\rho$-like part, whose corresponding consistent
MECs then serve as a basis for a consistent MEC for the given
potential except for a small modification of the short range part. 

The feasibility of this method by discretizing the continuous
superposition into a finite number of Yukawa functions has been 
demonstrated first for the Paris potential for which a consistent 
$r$-space MEC exists already. For a given grid of $N$ masses the
corresponding coefficients of the Yukawa functions are uniquely
determined by a properly chosen grid of $N$ radial points and involve
a simple matrix inversion only. It turned out that the convergence
with the number of terms is very rapid, and that with $N=16$ one
obtains an excellent representation of the potential. The same was
found for the more recent Argonne $V_{18}$ potential.

The resulting consistent MEC, represented by a corresponding
superposition of $\pi$- and $\rho$-like MECs, has then be checked 
by evaluating the inclusive cross section of deuteron 
electrodisintegration near threshold. For the associated 
observable, the inelastic transverse form factor, which is
sensitive to MEC, we found for the Paris potential excellent agreement
with previous evaluations and for both potentials a very rapid convergence 
with the number of terms. Thus the present method will easily allow one
to implement a consistent MEC into an $r$-space calculation using 
phenomenological $NN$-potentials. 

Finally we would like to emphasize that, although consistency of the MEC with 
the potential is achieved, one has to be aware of the fact, that this 
MEC is by no means unique. We have already alluded to a certain arbitrariness
in separating a short-range part in order to eliminate an otherwise appearing
$\delta$-function. Furthermore, there is in addition a freedom in the
spin-operator structure as has been pointed out already in~\cite{BuL85}. 
Only the longest range part of the $\pi$-like exchange current, namely 
the genuine $\pi$-MEC is on safe grounds.

\newpage
\begin{table}
\caption{Parameter value $b$ for the radial grid, scale parameter $s$
for the mass grid of Gauss points and relative mean deviation for Paris 
potential as function of the number of Gauss points $N$.}
\begin{center}
\begin{tabular}{cccccc}
$N$ & 12 & 14 & 16 & 18 & 20\\
$b$ & 0.42 & 0.34 & 0.29 & 0.245 & 0.21\\
$s$ [fm$^{-1}$] & 1.64 & 2.20 & 2.67 & 3.09 & 3.48\\
$\Delta(V_C^{(Y,N)})/|\bar V_C|$ & $0.25\cdot 10^{-4}$ & $0.21\cdot 10^{-5}$ & 
$0.54\cdot 10^{-6}$ & $0.72\cdot 10^{-7}$ & $0.14\cdot 10^{-7}$\\
$\Delta(V_T^{(Y,N)})/|\bar V_T|$ & $0.20\cdot 10^{-3}$ & $0.35\cdot 10^{-4}$ & 
$0.72\cdot 10^{-5}$ & $0.65\cdot 10^{-6}$ & $0.12\cdot 10^{-6}$\\
\end{tabular}
\label{tab1}
\end{center}
\end{table}

\begin{table}
\caption{Scale parameter $s$ and relative mean deviation for Argonne $V_{18}$ 
potential as function of the number of Gauss points $N$.}
\begin{center}
\begin{tabular}{cccccc}
$N$ & 12 & 14 & 16 & 18 & 20\\
$s$ [fm$^{-1}$] & 2.19 & 2.43 & 2.75 & 3.04 & 3.235\\
$\Delta(V_C^{(Y,N)})/|\bar V_C|$ & $0.21\cdot 10^{-3}$ & $0.69\cdot 10^{-4}$ & 
$0.58\cdot 10^{-4}$ & $0.45\cdot 10^{-4}$ & $0.36\cdot 10^{-4}$\\
$\Delta(V_T^{(Y,N)})/|\bar V_T|$ & $0.61\cdot 10^{-3}$ & $0.24\cdot 10^{-3}$ & 
$0.11\cdot 10^{-3}$ & $0.31\cdot 10^{-4}$ & $0.98\cdot 10^{-5}$\\
\end{tabular}
\end{center}
\label{tab2}
\end{table}

\begin{references}
\bibitem{Mac89}
R. Machleidt, Adv. Nucl. Phys. {\bf 19}, 189 (1989).

\bibitem{RiG97}
F. Ritz, H. G\"oller, T. Wilbois, and H.\ Arenh\"ovel, 
Phys. Rev. C {\bf 55}, 2214 (1997). 

\bibitem{ChR71}
M. Chemtob, M. Rho, Nucl. Phys. {\bf A163}, 1 (1971). 

\bibitem{Mat89}
J.F. Mathiot, Phys. Rep. {\bf 173}, 63 (1989).

\bibitem{Ris89}
D.O.\ Riska, Phys.Rep. {\bf 181}, 207 (1989). 

\bibitem{Are99}
H.\ Arenh\"ovel, Few-Body Syst. {\bf 26}, 43 (1999).

\bibitem{BuL85}
A.\ Buchmann, W.\ Leidemann, and H.\ Arenh\"ovel, Nucl.\ Phys.\ {\bf A443},
726 (1985).

\bibitem{Ris85}
D.O.\ Riska, Phys.\ Scrip.\ {\bf 31}, 471 (1985).

\bibitem{LaL80}
M.\ Lacombe, B.\ Loiseau, J.M.\ Richard, and R.\ Vinh Mau, 
Phys.\ Rev.\ C {\bf 21}, 861 (1980).

\bibitem{WiS95}
R.B. Wiringa, V.G. Stoks, and R. Schiavilla, Phys. Rev. C {\bf 52}, 38 (1995).

\bibitem{HoR73}
J. Hockert, D.O. Riska, M. Gari, and A. Huffmann, Nucl. Phys. 
{\bf A217}, 14 (1973).

\bibitem{LoF75}
J.A. Lock and L.L. Foldy, Ann. Phys. (N.Y.) {\bf 93}, 276 (1975).

\bibitem{FaA76}
W. Fabian and H. Arenh\"ovel, Nucl.\ Phys.\ {\bf A258}, 461 (1976).

\bibitem{MoR76}
B. Mosconi and P. Ricci, Nuovo Cimento A {\bf 36}, 67 (1976).

\bibitem{BeJ81}
M.\ Bernheim, E.\ Jans, J.\ Mougey, D.\ Royer, D.\ Tarnowski,
S.\ Turck-Chieze, I.\ Sick, G.P.\ Capitani, E.\ De Sanctis,
and S.\ Frullani,
Phys.\ Rev.\ Lett.\ {\bf 46}, 402 (1981).

\bibitem{Auf85} 
S.\ Auffret {\em et al.},
Phys.\ Rev.\ Lett.\ {\bf 55}, 1362 (1985).

\bibitem{Arn90}
R.G.\ Arnold {\em et al.},
Phys.\ Rev.\ C {\bf 42}, R1 (1990).

\bibitem{BeW94}
G.\ Beck, T.\ Wilbois, and H.\ Arenh\"ovel,
Few-Body Syst.\ {\bf 17}, 91 (1994).

\end{references}
\end{document}